\documentclass[conference, 10pt]{IEEEtran}
\IEEEoverridecommandlockouts

\usepackage{cite}
\usepackage{amsmath,amssymb,amsfonts}
\usepackage{algorithm}
\usepackage{algorithmic}
\usepackage{graphicx}
\usepackage{epstopdf}
\usepackage{mathrsfs}
\usepackage{verbatim}
\usepackage{textcomp}
\usepackage{xcolor}
 
\DeclareMathOperator{\rect}{rect}

\def\BibTeX{{\rm B\kern-.05em{\sc i\kern-.025em b}\kern-.08em
    T\kern-.1667em\lower.7ex\hbox{E}\kern-.125emX}}
\begin{document}
\title{Wideband MIMO Beampattern Synthesis using Adaptive Frequency Modulated Waveforms
\thanks{This research was funded by the Naval Undersea Warfare Center's In-House Laboratory Independent Research (ILIR) program from the Office of Naval Research (ONR) under N0001424WX00177.}
}

\author{\IEEEauthorblockN{David A. Hague}
\IEEEauthorblockA{\textit{Naval Undersea Warfare Center} \\
1176 Howell St., Newport, RI, USA \\
david.a.hague.civ@us.navy.mil}
}

\maketitle

\begin{abstract}
Multiple-Input Multiple-Output (MIMO) arrays transmit unique waveforms on each of their elements increasing the degrees of freedom available to synthesize novel transmit beampatterns.  The wideband MIMO beampattern shape is determined by the structure of the MIMO Cross Spectral Density Matrix (CSDM) whose elements are the cross-spectra between each waveform in the set as a function of frequency.  This paper demonstrates a model for synthesizing wideband MIMO beampatterns using Multi-Tone Sinusoidal Frequency Modulated (MTSFM) waveform sets.  The MTSFM waveform's instantaneous phase is a finite Fourier series.  The Fourier coefficients are modified to synthesize constant modulus and spectrally compact waveforms whose CSDM closely approximates a desired wideband  MIMO transmit beampattern.  An optimization routine is formulated that synthesizes MTSFM waveform sets that possess the necessary CSDM structure and is demonstrated using an illustrative design example.
\end{abstract}


\section{Introduction}
Multiple-Input Multiple-Output (MIMO) arrays have seen extensive use in communications \cite{MIMO_Comms}, radar \cite{JianLiBookI}, and, more recently, in Integrated Sensing and Communications (ISAC) applications \cite{MIMO_ISAC}.  A coherent MIMO array is a generalization of the phased-array that transmits a unique waveform on each element.  This increases the degrees of freedom in the transmit array design which facilitates synthesizing novel transmit beampattern shapes.  The structure of the wideband beampattern shape is determined by the Cross-Spectral Density Matrix (CSDM), a frequency dependent matrix whose entries are the cross spectra between each waveform in the set across frequency.  The wideband MIMO beampattern synthesis problem is thus a multi-objective waveform optimization problem. The designer seeks a waveform set whose CSDM possesses the necessary structure across the operational band of frequencies to approximate a desired beampattern shape.  In addition to realizing a desired beampattern shape, the waveform sets should also possess a constant modulus and a compact spectral shape.  The constant modulus property ensures the waveform is amenable to efficient transmission on practical power amplifiers.  Spectrally compact waveforms densely concentrate their energy in a compact operational band of frequencies for more efficient operation in spectrally crowded environments \cite{Aubry_Spectra_I, Tang_Spectra}, a consideration that is of increasing importance to the radar community \cite{Spectral_Coexistence}.

There is an extensive and rich literature on designing waveform sets to approximate a desired narrowband MIMO transmit beampattern and a complete literature review is not possible here.  The reader is referred to \cite{MIMO_SanAntonio, jianLiII, MIMO_LI, MIMO_Li_CAN, Fourier_MIMO, PalomarI} for an overview of these narrowband design methods.  These efforts focused on waveform design scenarios with a small fractional bandwidth.  The fractional bandwidth $\gamma=\Delta f/(2f_c)$ is the ratio of a waveform's swept bandwidth $\Delta f$ over twice the carrier frequency $f_c$ where $0 < \gamma < 1.0$.  When the fractional bandwidth is small (i.e, $\gamma \ll 0.1$), the waveforms are considered narrowband and the MIMO beampattern is realized by shaping the waveform correlation matrix whose entries are the inner products between each waveform in the set \cite{MIMO_SanAntonio}.  Narrowband MIMO beampattern synthesis is an active research topic \cite{MIMO_Radar_Manifold} and is finding use in ISAC systems \cite{MIMO_ISAC_ICASSP_2024}.

When $\gamma \geq 0.1$, the waveforms are considered to be wideband and the MIMO beampattern is dependent on the sum of the entries of the waveform set's CSDM across the band of frequencies where the waveforms have substantial energy.  While there are considerably less efforts on wideband MIMO beampattern synthesis, there are still several notable results in the literature.  Work by SanAntonio et. al \cite{SanAntonio_MIMO_Wideband, SanAntonio_MIMO_Wideband_Journal} appears to have first formulated the wideband MIMO problem.  Later efforts by \cite{Wideband_MIMO_Systems_Stoica, Wideband_MIMO_Alg_ADMM, WideBand_MIMO_Majorization_ADMM_Spectral_Agility, MIMO_Wideband_Spectrally_Crowded} developed a suite of computationally efficient algorithms to synthesize waveform sets for wideband MIMO beampattern synthesis.  There are additionally a number of efforts that focused on ensuring the wideband MIMO beampattern's main beam width does not vary substantially across frequency thus simplifiying the waveform design process \cite{Wideband_MIMO_Frequency_Invariance, Wideband_MIMO_Shannon_Conference}.  All of the aforementioned wideband MIMO efforts focused on designs where $\gamma = 0.1$, on the cusp of the wideband regime.  However, work by \cite{Wideband_MIMO_Radar_Conference, Wideband_MIMO_Shannon} explored ``ultra-wideband'' MIMO radar waveform designs with $\gamma \approx 0.25$.  

The majority of both narrowband and wideband MIMO beampattern synthesis efforts share a common design challenge in developing constant modulus and spectrally compact waveform sets.  These design characteristics often take the form of constraints within the respective waveform optimization problem \cite{RangaswamyI} thus increasing its complexity.  This paper investigates applying the Multi-Tone Sinusoidal Frequency Modulated (MTSFM) waveform model \cite{Hague_AES} to design constant modulus and spectrally compact waveform sets whose CSDM closely approximates a desired wideband MIMO transmit beampattern.  The MTSFM waveform possesses an instantaneous phase that is represented as a finite Fourier series.  In previous efforts, these Fourier coefficients were modified to synthesize MTSFM waveform sets with a desired narrowband MIMO transmit beampattern shape \cite{Hague_Asilomar_2024, Hague_IEEE_SPL_2025}.  In this work, the Fourier coefficients will be modified to synthesize MIMO waveform sets with the desired CSDM structure.  Additionally, the MTSFM waveform naturally possesses the constant modulus and spectral compactness properties necessary for efficient transmission on practical radar transmitter electronics \cite{Hague_AES}.  These qualities make the MTSFM waveform an intriguing option for wideband MIMO radar systems.  A simple optimization method is formulated to design constant modulus and spectrally compact wideband MIMO MTSFM waveform sets and is then demonstrated via an illustrative design example.  


\section{wideband MIMO Waveform Signal Model}
\label{sec:sigModel}
This section describes the wideband MIMO waveform signal model and the MTSFM waveform.

\subsection{The MIMO Waveform Set}
The MIMO model used in this paper assumes a Uniform Linear Array (ULA) composed of $M$ elements with inter-element spacing $d=\lambda_d/2$ where $\lambda_d = c/f_d$ is the wavelength corresponding to the array's design frequency $f_d$ and $c$ is the speed of propagation in the medium.  Each element of the MIMO ULA transmits a unique waveform $s_m\left(t\right)$ where $m=1,\dots,M$.  Each of the $M$ waveforms in the waveform set are of duration $T$ and are expressed as
\begin{equation}
s_m\left(t\right) = \dfrac{\rect\left(t/T\right)}{\sqrt{MT}}e^{j\varphi_m\left(t\right)}e^{j2\pi f_c t},~-T/2\leq t\leq T/2
\label{eq:waveformSet}
\end{equation}
where $\varphi_m\left(t\right)$ is the $m^{\text{th}}$ waveform's instantaneous phase, $f_c$ is the carrier frequency of each waveform, and the $1/\sqrt{MT}$ term normalizes each waveform to have a total energy of $1/M$.  The spectra of each waveform in the set is denoted as $S_m\left(f\right)$.  The waveform's frequency modulation function is expressed as $\dot{\varphi}_m\left(t\right)/\left(2\pi\right)$ where $\dot{\varphi}_m\left(t\right)$ is the first time derivative of the instantaneous phase $\varphi_m\left(t\right)$.


\subsection{The Wideband MIMO Beampattern}
\label{subsec:ulsModel}
The wideband MIMO transmit beampattern is the normalized power density at angle $u=\sin\theta$ and denoted as $P\left(u\right)$ expressed as 
\begin{equation}
P\left(u\right) = \int_{-\infty}^{\infty}\mathbf{a}^{\text{H}}\left(u, f\right) \mathbf{S}\left(f\right) \mathbf{a}\left(u, f\right)df
\label{eq:mimoBeamPattern}
\end{equation}
where $P\left(u\right) \geq 0$ and $\mathbf{a}\left(u, f\right) \in \mathbb{C}^{M\times 1}$ is the frequency dependent transmit array steering vector at frequency $f$ expressed as \cite{SanAntonio_MIMO_Wideband}
\begin{equation}
\mathbf{a}\left(u, f\right) = e^{j2\pi\left(m-1\right)\left(d/\lambda\right)u} = e^{j\pi\left(m-1\right)\left(f/f_d\right)u}.
\label{eq:mimoSteerVec}
\end{equation} 
The matrix $\mathbf{S}\left(f\right) \in \mathbb{C}^{M\times M}$ is the MIMO CSDM of the waveform set with the entries $S_{m,m'}\left(f\right)=S_m\left(f\right)S_{m'}^*\left(f\right)$ representing the cross-spectral density between any two waveforms in the waveform set.  The expressions for the wideband MIMO beampattern \eqref{eq:mimoBeamPattern} and \eqref{eq:mimoSteerVec} are a more general model for the MIMO beampattern of which the narrowband equations are a special case.  When the waveform set is considered narrowband (i.e, $\gamma \ll 0.1$) and $f_c = f_d$, the spectra of each waveform is densely concentrated at the array's design frequency.  The array steering vector is then no longer frequency dependent and simplifies to 

\begin{equation}
\mathbf{a}\left(u\right) = e^{j\pi\left(m-1\right)u}.
\label{eq:steeringVecNarrow}
\end{equation}
The MIMO beampattern \eqref{eq:mimoBeamPattern} can then be expressed as
\begin{equation}
P\left(u\right) = \mathbf{a}^{\text{H}}\left(u\right)\left[\int_{-\infty}^{\infty}\mathbf{S}\left(f\right) df \right] \mathbf{a}\left(u\right)
\label{eq:wideMIMO}
\end{equation}
where the integral of each element of $\mathbf{S}\left(f\right)$ in \eqref{eq:wideMIMO} is simply 
\begin{equation}
\int_{-\infty}^{\infty}S_m\left(f\right) S_{m'}^*\left(f\right) df = \int_{-T/2}^{T/2} s_m\left(t\right) s_{m'}^*\left(t\right) dt
\label{eq:corrMat}
\end{equation}
which is the narrowband MIMO correlation matrix $\mathbf{R}= \langle s_m\left(t\right), s_{m'}^*\left(t\right)\rangle \in \mathbb{C}^{M\times M}$.


\subsection{The MTSFM Waveform Model}
\label{subsec:mimoMTSFM}
This paper utilizes the MTSFM waveform model in \cite{Hague_AES} whose phase modulation function is a finite Fourier series
\begin{equation}
\varphi_m\left(t\right) = \sum_{p=1}^P \alpha_{m,p} \sin \left(\dfrac{2\pi p t}{T}\right)
\label{eq:MTSFM}
\end{equation}
where $\alpha_{m,p}$ are the MTSFM waveform set's modulation indices.  The coefficients $\alpha_{m,0}$ are all assumed to be 0.  It is important to note that \eqref{eq:MTSFM} can also include cosine harmonics in its instantaneous phase.  However, for mathematical simplicty, this paper focuses solely on waveforms whose modulation functions possesses sine harmonics.   The modulation indices $\alpha_{m,p}$ are utilized as a discrete set of $MP$ design parameters that are modified to synthesize constant modulus and spectrally compact MTSFM waveform sets that closely approximate a desired wideband MIMO beampattern shape.  


\section{Designing Wideband MIMO Beampatterns using MTSFM Waveforms}
\label{sec:broadMimoMtsfm}
Direct computation of \eqref{eq:mimoBeamPattern} using the MTSFM spectra does not lead to a simple closed form solution to the wideband MIMO beampattern of a MTSFM waveform set.  However, as was noted in \cite{SanAntonio_MIMO_Wideband}, the wideband MIMO beampattern integral in \eqref{eq:mimoBeamPattern} can be well approximated as a sum over an appropriately chosen set of $N$ discrete frequencies $f_k$  
\begin{align} 
P\left(u\right) &\approx \sum_{k}P\left(u, f_k\right)
\label{eq:wideMimoDiscrete}
\end{align}
where  $P\left(u, f_k\right)=\mathbf{a}^{\text{H}}\left(u, f_k\right) \mathbf{S}\left(f_k\right) \mathbf{a}\left(u, f_k\right)$ is the discrete frequency and angle dependent power density and $f_k = f_c + k/T$. Each of the $M$ waveforms in the basebanded waveform set possess $N$ discrete time samples and are expressed as
\begin{equation}
s_m\left[n\right] = \sqrt{\dfrac{1}{MN}}\exp\left[j\sum_{p=1}^P \alpha_{m,p} \sin \left(\dfrac{2\pi p n}{N}\right)\right]
\label{eq:waveformSet}
\end{equation}
where $n=0,1,\dots,N-1$ and the $\sqrt{1/MN}$ term normalizes each waveform to have a total energy of $1/M$.  

As explained in \cite{Hague_AES, Hague_Asilomar_2024}, the MTSFM discrete time-series can be approximated as a Discrete Fourier Series (DFS) via the Jacobi-Anger expansion for Generalized Bessel Functions (GBFs) \cite{Dattoli}
\begin{equation}
s_m\left[n\right] \cong \sqrt{\dfrac{1}{MN}}\sum_{k=-N/2}^{N/2-1}\mathcal{J}_k^{1:P}\left(\{\alpha_{m,p}\}\right) e^{j\frac{2\pi k n}{N}}.
\label{eq:mtsfmJAE}
\end{equation}  
The $\mathcal{J}_k^{1:P}\left(\{\alpha_{m,p}\}\right) = \mathcal{J}_k\left(\alpha_{m,1}, \alpha_{m, 2},\dots, \alpha_{m,P}\right)$ term is the $k^{\text{th}}$ order $P$-dimensional cylindrical GBF \cite{Dattoli} and represents the DFS coefficients $S_m\left[k\right]$ of the $m^{\text{th}}$ waveform in the set.  Note that $\mathcal{J}_k^{1:P}\left(\{\alpha_{m,p}\}\right)$ is non-zero for all $k$ but is vanishingly small for $|k| \geq N/2$ if $C\|\alpha_{m, p}\|_2 \leq N/2$ for all $M$ waveforms in the set where $C$ is a constant \cite{Dattoli}.  

Each element in the CSDM $S_{m,m'}\left[k\right] = S_m\left[k\right]S^*_{m'}\left[k\right]$ is expressed in terms of the DFS coefficients $S_{m,m'}\left[k\right] = \mathcal{J}_k^{1:P}\left(\{\alpha_{m,p}\}\right) \mathcal{J}_k^{1:P}\left(\{\alpha_{m',p}\}\right)$ and is real valued since the cylindrical GBFs are real-valued.  This discretized version of the wideband MIMO beampattern in \eqref{eq:wideMimoDiscrete} with these GBF CSDM elements are a general expression for which the narrowband MIMO beampattern equations are again a special case.  For low fractional bandwidths (i.e, $\gamma \ll 0.1$), the array steering vectors $\mathbf{a}\left(u, f_k\right)$ are no longer frequency dependent and simplify to $\mathbf{a}\left(u\right)$ from \eqref{eq:steeringVecNarrow} which in turn simplifies \eqref{eq:wideMimoDiscrete} to 
\begin{equation}
P\left(u\right) = \mathbf{a}^{\text{H}}\left(u\right) \left[\sum_{k=-N/2}^{N/2-1}\mathbf{S}\left(f_k\right)\right]\mathbf{a}\left(u\right).
\label{eq:wideMimoDiscreteNarrow}
\end{equation}
Summation of $\mathbf{S}\left(f_k\right)$ for all $k$ results in the narrowband MIMO correlation matrix whose elements are $\sum_{k}S_{m,m'}\left[k\right]$.  Leveraging an identity for GBFs \cite{Dattoli}, $S_{m,m'}\left[k\right]$ approximately simplifies to
\begin{equation}
\sum_{k=-N/2}^{N/2-1}S_{m,m'}\left[k\right] \approx \mathcal{J}_0^{1:P}\left(\{\alpha_{m,p}-\alpha_{m',p}\}\right) 
\label{eq:gbfIdentity}
\end{equation}
which is the expression for each element in the correlation matrix of a MTSFM waveform set that was derived in \cite{Hague_Asilomar_2024}.

\subsection{A Waveform Set Analysis Example}
\label{subsec:analysisExample}
\begin{figure}[htbp]
\centering
\includegraphics[width=0.5\textwidth]{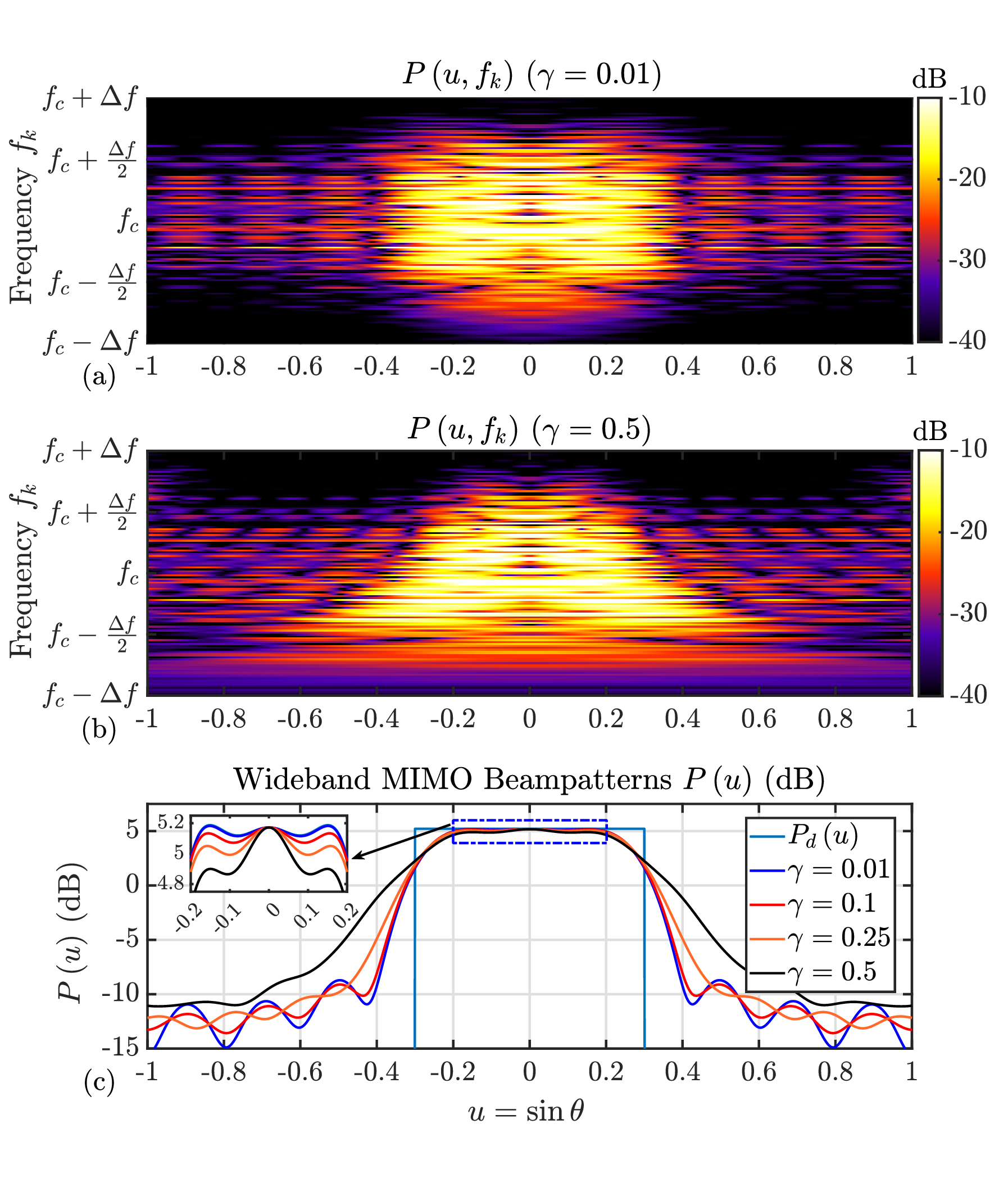}
\caption{Illustration of $P\left(u, f_k\right)$ for (a) $\gamma = 0.01$ and (b) for $\gamma = 0.5$ along with the resulting beampatterns $P\left(u\right)$ for a range of fractional bandwidths (c).  For increasingly wideband MIMO beampatterns, the low frequency beampattern components become gradually broader in $u$ resulting in a widening of the overall main beam width while also smearing the sidelobes and increasing their height.  The high frequency components become more concentrated near $u=0$ and increases the ripple levels within the main beam.}
\label{fig:MIMO_1}
\end{figure}

To fully demonstrate the model for designing MTSFM waveform sets for wideband MIMO, it is insightful to first consider a design example and analyze $P\left(u, f_k\right)$ across a range of fractional bandwidths.  This example considers a ULA with $M=10$ sensors using a MTSFM waveform set from \cite{Hague_Asilomar_2024}.  Each waveform in the set ($M=10$ waveforms) possesses $P=16$ modulation indices and a time-bandwidth product $T\Delta f = 64$.  The original ULA design frequency $f_d = f_c$ was chosen such that $\gamma = 0.01$ thus keeping the array and waveforms firmly in the narrowband regime.  The waveforms were optimized using the optimization routine developed in \cite{Hague_Asilomar_2024} that sought to provide a Minimum Mean-Square Error (MMSE) \cite{MIMO_SanAntonio} approximation to the desired MIMO beampattern shape $P_d\left(u\right)$ 
\begin{equation}  P_d\left(u\right) = \left\{
\begin{array}{ll}
	A, & |u=\sin\theta| \leq 0.3\\

      0, & owise\\
\end{array} 
\right.
\label{eq:region}   
\end{equation}
where A is a normalization constant that ensures the total power across $u$ equals the dimensionality of the array $M$ \cite{MIMO_SanAntonio}.  Figure \ref{fig:MIMO_1} shows the resulting MIMO beampatterns across a range of fractional bandwidths $\gamma$ resulting from evaluating \eqref{eq:wideMimoDiscrete}.  Also shown is $P\left(u, f_k\right)$ for the narrowband case ($\gamma = 0.01$) and a wideband case ($\gamma = 0.5$) across $N=128$ frequencies spanning $f_c-\Delta f \leq f_k < f_c+\Delta f$.  In order to implement the wideband MIMO beampattern for higher fractional bandwidths using MTSFM waveforms with a fixed bandwidth $\Delta f$, $f_d$ and $f_c$ must be lowered commensurate with $\gamma$.  In order to keep the same effective mainbeam resolution at these lower design frequencies, the aperture length must be increased by setting the inter-element spacing to 
\begin{equation}
d = \dfrac{\lambda_d}{2} = \dfrac{c}{2f_d} = \dfrac{c}{2\left(f_c+\Delta f\right)}
\label{eq:wideSpacing}
\end{equation}
which additionally avoids spatial aliasing.  

The first observation from Figure \ref{fig:MIMO_1} is that $P\left(u, f_k\right)$ is densely concentrated in the band of frequencies $f_c-\Delta f /2 \leq f_k < f_c + \Delta f/2$ for both when $\gamma=0.01$ and $\gamma=0.5$.  This is due to the compact spectral shapes of both MTSFM waveform sets and is a natural characteristic of the MTSFM waveform model \cite{Hague_AES, Hague_Asilomar_2024}.  Summing $P\left(u, f_k\right)$ across $f_k$ results in the wideband MIMO beampattern $P\left(u\right)$.  The beampattern that results from the $\gamma = 0.01$ configuration is indistinguishable from the narrowband MIMO beampattern formulation.  This is because the array steering vector $\mathbf{a}\left(u, f_k\right)$ is largely invariant for small $\gamma$ in accordance with the narrowband special case of the wideband model as shown in \eqref{eq:steeringVecNarrow}-\eqref{eq:corrMat}.  However, the beampatterns begin to distort with increasing $\gamma$ and there are two main reasons for this.  The first is that array steering vector $\mathbf{a}\left(u, f\right)$ in $P\left(u, f_k\right)$ for frequencies well below $f_d$ are increasingly spread out in $u$.  As can be seen in panel (c) of Figure \ref{fig:MIMO_1}, this results in a widening of the mainbeam transition width and a smearing and of the sidelobe levels which also increases their height.  The second reason is the high frequency components of $P\left(u, f_k\right)$ are increasingly concentrated near $u=0$.  This is due to $\mathbf{a}\left(u, f_k\right)$ being more narrowly concentrated in $u$ for frequencies well above $f_d$.  Summing these concentrated high frequency components of $P\left(u, f_k\right)$ across $f_k$ results in an increase in the ripple levels of the mainbeam response.  While this MTSFM waveform set largely met the narrowband MIMO beampattern design criteria, its resulting wideband MIMO beampatterns become progessively more perturbed with increasing $\gamma$.  When operating in the wideband regime, the waveform set must be re-optimized so that the resulting MIMO beampattern better meets the design criteria in \eqref{eq:region}.

\subsection{A Waveform Set Optimization Example}
\label{subsec:analysisExample}
Optimizing the MTSFM waveform set to possess the desired wideband MIMO beampattern shape in \eqref{eq:region} requires reformulating the MMSE optimization problem from \cite{MIMO_SanAntonio, Hague_Asilomar_2024} to the broadband regime as in \cite{SanAntonio_MIMO_Wideband}.  Formally, this problem for optimizing MTSFM waveform sets is defined as 
\begin{align}
\underset{\alpha_{m,p}}{\text{min}}\Biggl\{\Bigl\| P_d\left(u\right)& -\sum_k P\left(u, f_k;\alpha_{m,p}\right)\Bigr\|_2^2\Biggr\} \IEEEnonumber \\ \text{s.t.~} &\beta_{rms}^2\left(\{\alpha_{m,p}^{(i)}\}\right) \in \left(1\pm\delta\right)\beta_{rms}^2(\{\alpha_{m,p}^{\left(0\right)}\}) \IEEEnonumber \\ &\sum_k S_{m,m} = 1/M~\forall m,~ \mathbf{S}\left(f_k\right) \succeq 0 %
\label{eq:Problem_1}
\end{align}
where $\alpha_{m,p}^{(i)}$ are the MTSFM waveform set's coefficients at iteration $i$ and thus $\alpha_{m,p}^{\left(0\right)}$ are the initial coefficients passed to the optimization routine.  The quantity $\beta_{rms}^2$ is the Root-Mean Square (RMS) bandwidth \cite{Cohen} for each waveform in the set and is expressed as $\beta_{rms}^2(\{\alpha_{m,p}\})=\sum_p p^2\alpha_{m,p}^2/2$ \cite{Dattoli, Hague_AES}.
The parameter $\delta$ is a user-defined unitless constant that bounds the RMS bandwidth.  This ensures that the bandwidths of each waveform in the optimized set do not substantially deviate from the bandwidths of each waveform in the initial set.  The second constraint ensures that each waveform preserves its total energy.  The last constraint  ensures that the CSDM at each frequency is positive semi-definite.  The problem posed in \eqref{eq:Problem_1} is numerically solved using the MATLAB$\textsuperscript{TM}$ function \emph{fmincon()} \cite{MATLAB}.  This function utilizes an interior point algorithm to find a local minimum to the waveform optimization problem in \eqref{eq:Problem_1} that also accomodates the nonlinear constraints.  It's important to note that this routine is not streamlined for computational efficiency and the development of algorithms towards this end will be the topic of a future paper.  %

\begin{figure}[htbp]
\centering
\includegraphics[width=0.5\textwidth]{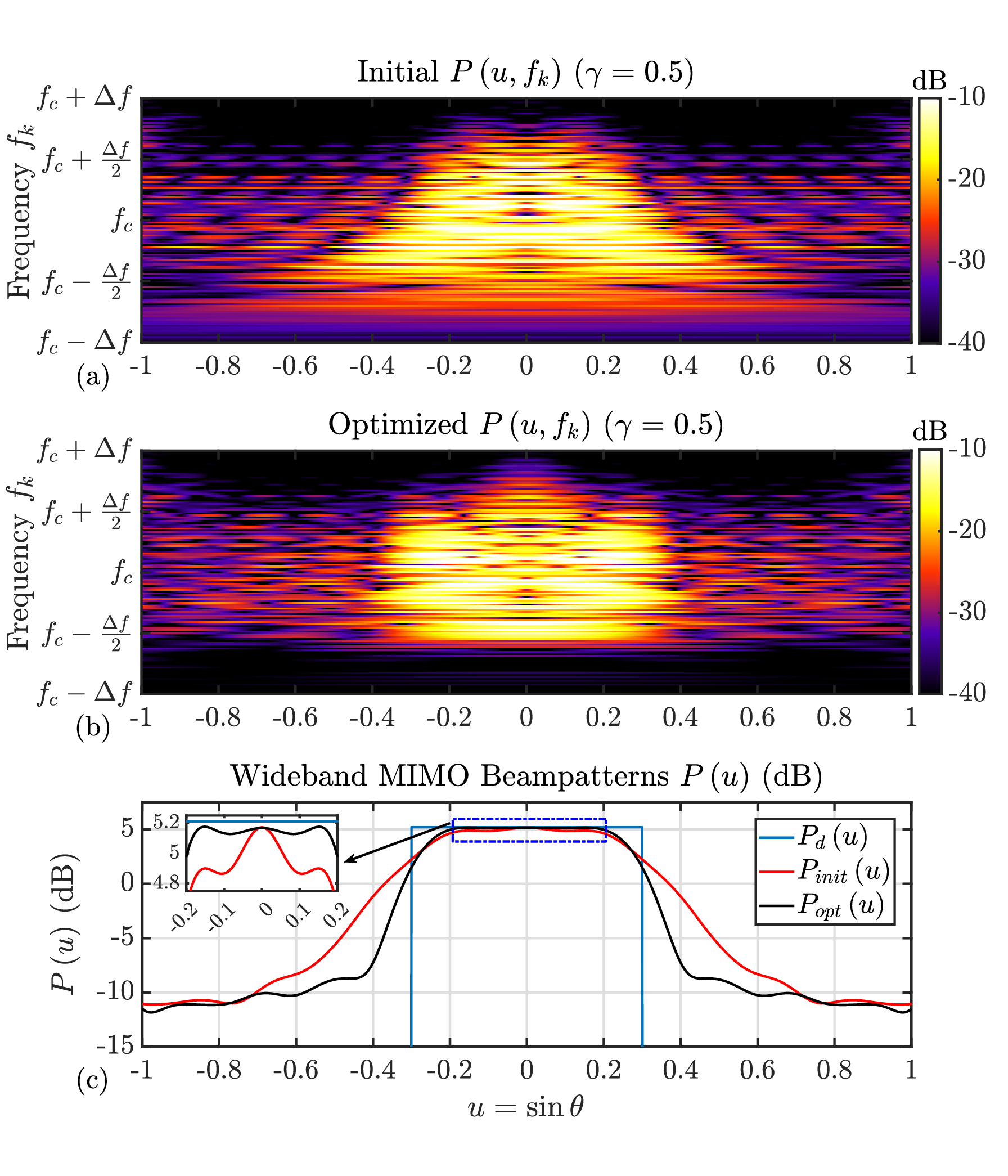}
\caption{$P\left(u, f_k\right)$ for (a) the initial MTSFM waveform set and (b) the optimized MTSFM waveform set using \eqref{eq:Problem_1} along with their corresponding wideband MIMO beampatterns $P\left(u\right)$ (c).  The optimized MTSFM waveform set more closely meets the desired beampattern criteria \eqref{eq:region} than the initial waveform set.}
\label{fig:MIMO_2}
\end{figure}

Figure \ref{fig:MIMO_2} shows $P\left(u, f_k\right)$ for the initial and optimized MTSFM waveform sets as well as their wideband MIMO beampattern shapes.  For this example, the initial coefficients $\alpha_{p,k}^{\left(0\right)}$ passed to the routine were from the MTSFM waveform set that produced the $P\left(u, f_k\right)$ seen in panel (b) of Figure \ref{fig:MIMO_1} with $\delta = 0.1$.  From the figure, it is clear that the resulting optimized waveform set's wideband MIMO beampattern more closely resembles the desired beampattern shape than the initial waveform set.  The optimized beampattern possesses a narrower transition between the main beam and sidelobe regions.  Additionally, the sidelobe heights were reduced as was the ripple within the mainbeam.  The reasons for this improvement in the beampattern shape can be gleaned from inspecting $P\left(u, f_k\right)$ of the initial and optimized waveform sets.  Recall that the low frequency content of $P\left(u, f_k\right)$ contributed to the wider mainbeam width and higher smeared sidelobes while the high frequency content contributed to the increase in ripple within the mainbeam.  The low frequency content in $P\left(u, f_k\right)$ of the optimized MTSFM waveform set is substantially reduced compared to that of the initial waveform set.  There is also much less variation in the width at each frequency component in $P\left(u, f_k\right)$ in the optimized waveform set's mainbeam transition width.  This loosely resembles some of the frequency invariant beampattern designs of \cite{Wideband_MIMO_Frequency_Invariance, Wideband_MIMO_Shannon_Conference}.  These two characteristics combine to produce the narrower mainbeam and lower overall sidelobes.  Additionally, the high frequency content of $P\left(u, f_k\right)$ in the optimized waveform set is also noticeably reduced.  This directly translated to less ripple in the mainbeam region of $P\left(u\right)$   compared to the initial waveform set.  
\section{Conclusion}
\label{sec:Conclusion}
This paper presents a first attempt at optimizing the wideband MIMO beampattern shape of MTSFM waveform sets operating in the very large fractional bandwidth regime (i.e, $\gamma = 0.5$).  The wideband MIMO beampattern of a MTSFM waveform set is well characterized by its CSDM which has an exact closed form in terms of cylindrical GBFs of the first kind.  This closed form directly informs the distribution of transmit power across angle and frequency $P\left(u, f_k\right)$ and consequently the final beampattern shape $P\left(u\right)$.  This model is then used to formulate an optimization routine that synthesizes novel MTSFM waveform sets whose wideband MIMO beampattern approximates a desired shape.  The resulting waveform sets evenly distribute transmit power across the desired mainbeam region and operational band of frequencies.  The two primary follow on efforts to this work are (1) the development of more computationally efficient algorithms for this problem and (2) exploring methods to directly shape $P\left(u, f_k\right)$ over distinct regions in angle and frequency simultaneously much like the efforts in \cite{Wideband_MIMO_Shannon}.


\bibliographystyle{IEEEtran}
\bibliography{Hague_EUSIPCO_2026}

\end{document}